\documentclass[entropy,article,accept,moreauthors,pdftex]{Definitions/mdpi} 

\usepackage[percent]{overpic}
\usepackage{color}

\firstpage{1} 
\pubvolume{1}
\issuenum{1}
\articlenumber{0}
\pubyear{2021}
\copyrightyear{2021}
\externaleditor{Academic Editor:Andrei Khrennikov} 
\datereceived{07 April 2021} 
\dateaccepted{12 May 2021} 
\datepublished{} 
\hreflink{https://doi.org/} 

\Title{Characterization of a Two-Photon Quantum Battery: Initial Conditions, Stability and Work Extraction}
\TitleCitation{Characterization of a Two-Photon Quantum Battery: Initial Conditions, Stability and Work Extraction}

\Author{Anna Delmonte $^{1}$, Alba Crescente $^{1,2}$, Matteo Carrega $^{2}$, Dario Ferraro $^{1,2,}$* and Maura Sassetti $^{1,2}$}
\AuthorCitation{Delmonte, A.; Crescente, A.; Carrega, M.; Ferraro, D.; Sassetti, M.}

\address{%
$^{1}$ \quad Dipartimento di Fisica, Universit\`a di Genova, Via Dodecaneso 33, 16146 Genova, Italy; anna.delmonts@gmail.com (A.D.); crescente@fisica.unige.it (A.C.); sassetti@fisica.unige.it (M.S.) \\
$^{2}$ \quad SPIN-CNR, Via Dodecaneso 33, 16146 Genova, Italy; matteo.carrega@spin.cnr.it
}

\corres{Correspondence: ferraro@fisica.unige.it} 

\abstract{We consider a quantum battery that is based on a two-level system coupled with a cavity radiation by means of a two-photon interaction. Various figures of merit, such as stored energy, average charging power, energy fluctuations, and extractable work are investigated, considering, as possible initial conditions for the cavity, a Fock state, a coherent state, and a squeezed state. We show that the first state leads to better performances for the battery. However, a coherent state with the same average number of photons, even if it is affected by stronger fluctuations in the stored energy, results in quite interesting performance, in particular since it allows for almost completely extracting the stored energy as usable work at short enough times.} 


\begin{document}
\section{Introduction}
Quantum thermodynamics is a growing field of research that aims at extending concepts, such as heat and work to the realm of quantum physics~\cite{Campisi11, Levy12, Pekola15, Vinjanampathy16, Benenti17, Campisi17, Thermodynamics18, Carrega19, Porta18}. In this framework, several conventional assumptions of classical thermodynamics should be reconsidered. In particular, one needs to depart from the concept of thermodynamic equilibrium to properly include quantum coherences or possible 
external drives acting on individual quantum systems. In recent years, a very active branch of this new field of research focused on the study of energy transfer and storage in quantum devices leading to the concept of Quantum Battery (QB)~\cite{Alicki13, Thermodynamics18, Bhattacharjee20}. 

Various strategies have been developed in order to exploit quantum features to outperform classical devices, in particular enhancing the charging power (the energy that is stored in a given time interval) and the extractable work~\cite{Alicki13, Hovhannisyan13, Binder15, Moraes20, Santos20, Bai20, GPintos20, Mitchison20, Caravelli21, Peng21, Liu21, Hu21}. Theoretical proposals for possible actual implementations of QBs are often based on engineered two-level systems (TLSs) such as trapped ions~\cite{Cirac95, Bruzewicz19, Georgescu20}, superconducting qubits~\cite{Devoret13} or quantum dot in semiconductors~\cite{Singha11}. These platforms, which are usually exploited for qubit implementations, can be used as QB and can be coupled to chargers of a different nature able to coherently transfer energy into them. The proposed charging mechanisms range from classical external drives~\cite{Zhang19, Crescente20, Chen20, Yang20} to a proper control of the interaction in an array of TLSs~\cite{Ziani13, Le20, Rosa20, Rossini20, Ghosh21}. 

In this respect, promising platforms are devices where a collection of TLSs is coupled to a monochromatic cavity radiation in the same spirit of what has been realized in cavity and circuit quantum electrodynamics~\cite{Haroche_Book, Schoelkopf08}. Indeed, it has been shown that, for such kind of QBs, an enhancement of the averaged charging power, which scales as $\sqrt{M}$, with $M$ being the total number of TLSs, can be achieved, due to the collective interaction among the TLSs~\cite{Ferraro18, Andolina19, Andolina19b, Ferraro19}. Interestingly, a similar scaling, compatible with theoretical predictions, has been reported in a very recent experiment~\cite{Quach20}, demonstrating, for the first time, the collective quantum advantage first discussed in Ref. \cite{Binder15}.

In addition, recent theoretical proposals have shown that it is possible to engineer, in both trapped ions~\cite{Felicetti15} and flux qubit systems~\cite{Felicetti18}, the suppression of the conventional dipole coupling, linear in the quantum field radiation, in order to access the peculiar 
phenomenologies, such as super-radiance and spectral collapse, which are associated 
to the two-photon coupling (quadratic in the radiation field)~\cite{Emary02, Dolya09, Chen12, Peng17, Garbe17, Garbe20}.
These ideas have also recently been considered in the context of QBs by some of us, showing that two-photon coupling, once realized, can lead to an 
even greater (proportional to $M$ instead of $\sqrt{M}$) enhancement of the averaged charging power~\cite{Crescente20b}. This analysis has been carried out considering as initial state for the cavity radiation a Fock state where the number of photons is exactly the double of the number of TLSs in such a way that, at resonance, all the radiation energy is transferred to the QB (charging) and back (discharging). However, this form for the initial state is quite ad hoc and not so easy to be implemented experimentally. According to this, it is relevant to investigate also other possible initial states for the cavity radiation, such as coherent states and squeezed states, which are easier to be realized from an experimental point of view~\cite{Haroche_Book, Fink09}.

In the present paper, we address this aspect to achieve a full characterization of the two-photon based charging mechanism. For the sake of clarity, we will consider the case of a single TLS as the QB to be charged, avoiding other competing mechanisms that are based on collective behaviour \cite{Le20, Zhang19}. We will also adopt the so-called rotating-wave approximation (RWA), in the same spirit of Ref.~\cite{Andolina18} for the single-photon case, which allows for a simple analytical treatment.
We will describe the charging of a QB investigating different figures of merit. In particular, we will study the energy storage and associated charging time. Moreover, we will consider the average charging power and the fluctuations of the stored energy~\cite{Crescente20, Crescente20b, Friis18, Caravelli20}. Finally, a discussion regarding the maximum extractable work, namely the maximum fraction of the energy stored into the battery, which can be actually extracted and used to further purposes, will be addressed. This quantity, which is known in the literature as ergotropy \cite{Allahverdyan04}, in general does not coincide with the total energy that is stored in the QB due to quantum correlations \cite{Andolina19, Santos20, Caravelli21}. This analysis has already been carried out for a conventional single-photon coupling \cite{Andolina19} identifying the Fock and coherent state as the best choices for extracting almost all of the stored energy, but is still lacking for a two-photon coupling.   

All of these different aspects will be analyzed while taking various initial conditions for the cavity states into account. In particular, we will consider 
a Fock state, a coherent state, and a squeezed state with the same average 
number of photons as representative of the more conventional quantum state for the cavity radiation. Our study confirms that a properly designed Fock state is optimal for the functioning of the QB leading to maximal stored energy, short charging time, and a good amount of extractable work. However, interesting performances can also be found for a coherent state with the same averaged number of photons. In particular, this state shows an important ratio between extractable work and stored energy at short enough times. An opposite behaviour is observed for a squeezed state whose performances are all around very poor. The present analysis could give relevant hints for actual experimental implementations of these kind of QBs. 

The paper is organized, as follows. In Section~\ref{sec:model}, we introduce the model of QB, where a single TLS is coupled to a unique cavity mode via a two-photon interaction. Here, we consider different initial states for the cavity such as a Fock state, a coherent state and a squeezed state. Section~\ref{sec:characterization} is devoted to the characterization of the QB for these three different cavity states. In particular, we study the charging of the QB when considering the stored energy and corresponding charging times, as well as the related charging power. Moreover, we investigate the energy fluctuations and maximum work that can be extracted from the QB. All of these figures of merit are relevant in view of future development of this kind of QBs. In Section~\ref{sec:conclusions}, we draw the conclusions of our work.


\section{Model \label{sec:model}}
We consider a QB described as a single TLS coupled to the radiation of a cavity through a two-photon coupling (quadratic in the electron field) with matter-radiation coupling $\lambda$. This simple single cell can be replicated $M$ times to form a QB that works in a parallel charging configuration~\cite{Andolina18, Ferraro18}. The theoretical possibility of engineering such a kind of coupling in realistic devices has been recently discussed for trapped ions subject to bichromatic driving~\cite{Felicetti15} and for flux qubit coupled with a symmetric dc superconducting quantum 
interference device (SQUID)~\cite{Felicetti18}. The corresponding Hamiltonian can be written as 

\begin{equation}
\label{Hgen} 
\hat{\mathcal{H}}_{\rm 2ph}=\frac{\omega_a}{2}\hat{\sigma}_{z}+ \omega_c\hat{a}^\dag\hat{a}+\theta(t)\lambda\left(\hat{a}+\hat{a}^{\dagger}\right)^{2}\hat{\sigma}_{x}, 
\end{equation}
where $\hat{a}$ ($\hat{a}^\dag$) are the annihilation (creation) operators 
of the cavity radiation, $\hat{\sigma}_x,\hat{\sigma}_z$ are the Pauli matrices, $\omega_{a}$ the level spacing of the TLS, and $\omega_{c}$ the frequency of the radiation in the cavity. We underline that the TLS and cavity do not interact at $t<0$ and we assume that the 
two-photon interaction is switched on at time $t\geq0$, as indicated by the $\theta(t)$ step function. According to this, it is possible to transfer energy from the cavity to the TLS and observe the charging of the QB.

\begin{figure}[H]
\includegraphics[scale=0.55]{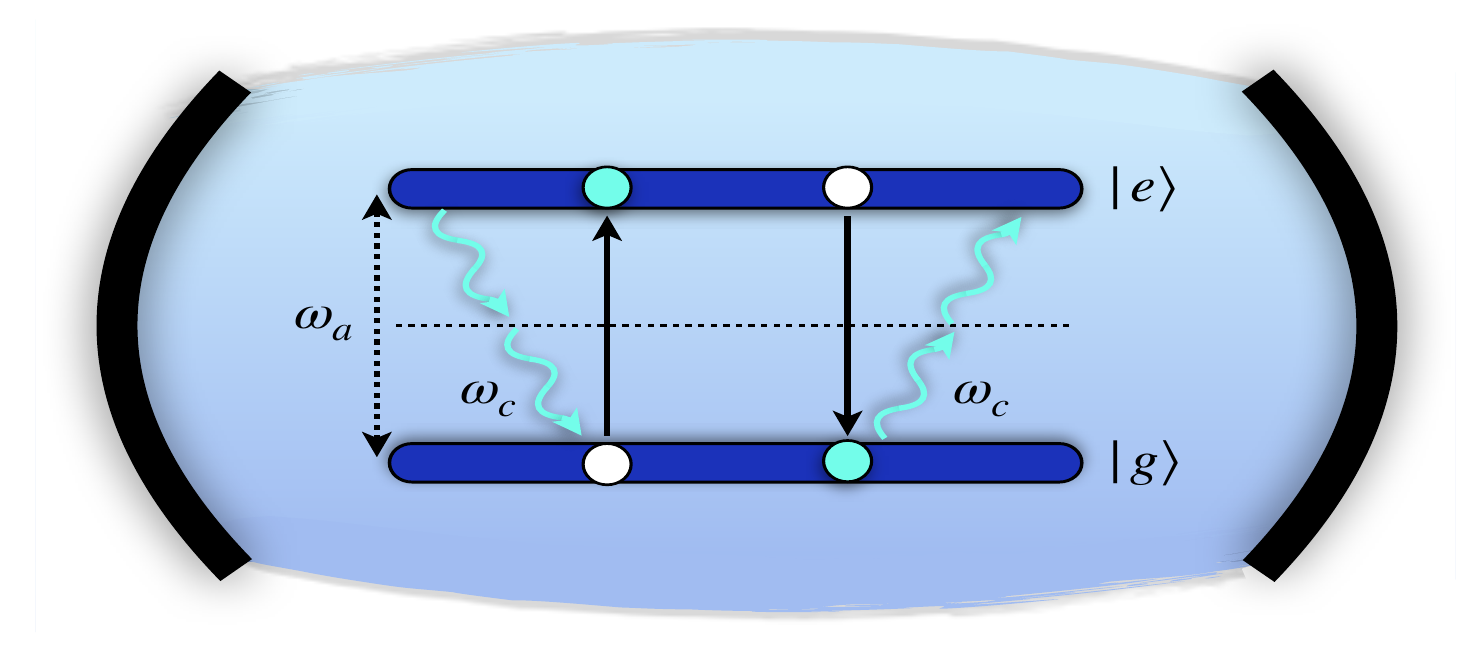}
\caption{Schematic representation of a QB where a TLS with level spacing $\omega_a$ between its ground ($|g\rangle$) and excited  ($|e\rangle$) state interacts with a single cavity mode of frequency $\omega_c$ via a 
two-photon coupling in the resonant regime $\omega_a=2\omega_c$.}
\label{Fig1}
\end{figure}

In the limit where the matter-radiation coupling $\lambda$ satisfies $\lambda\ll \omega_{a}, \omega_{c}$, one can consider the RWA of Equation~(\ref{Hgen}) and describe the system in terms of the two-photon Jaynes--Cummings model~\cite{JC63, Bartzis91, Crescente20b} with Hamiltonian being given by

\begin{equation}
\label{Htot} 
\hat{\mathcal{H}}= \frac{\omega_a}{2}\hat{\sigma}_z+\omega_c\hat{a}^\dag\hat{a}+\theta(t)\lambda(\hat{a}^{\dag^2}\hat{\sigma}_{-}+\hat{a}^2\hat{\sigma}_{+}), 
\end{equation}
where the first term is associated with QB, the second with the cavity micro-wave radiation used as charger, and the last one is the two-photon interaction term in the RWA~\cite{Felicetti15, Felicetti18} and with 
\begin{equation}
\hat{\sigma}_\pm=\frac{\hat{\sigma}_{x}\pm i\hat{\sigma}_{y}}{2},
\end{equation}
\noindent where $\hat{\sigma}_y$ is one of the Pauli matrices.

In the following, we will focus on the resonant case $\omega_{a}=2\omega_{c}$ (see  Figure~\ref{Fig1}), where all of the energies associated to the radiation can be transferred to the TLS and viceversa~\cite{Schleich_book}. This condition guarantees an optimal functioning of the QB, which is not achieved in other configurations of the device~\cite{Crescente20b}.

 The initial condition at time $t=0$ for the total system is chosen as a tensor product state
\begin{equation}\label{initial}
    |\psi(0)\rangle =|g\rangle \otimes \sum_n \alpha_{n} |n\rangle,
\end{equation}

\noindent where the first factor represents the ground state of the TLS, while the second characterizes the charger with $|n \rangle$, a state with $n$ photons in the cavity. We will consider, as paradigmatic examples, a Fock (fixed number of photons), a coherent, and a squeezed state for the cavity.  
Thus, we introduce the probability amplitudes $\alpha_{n}$, such that

\begin{eqnarray}
\label{alphaF} \alpha^{(F)}_n&=&\delta_{n,N},  \\
\label{alphaC} \alpha^{(C)}_n&=&e^{-\frac{N}{2}}\frac{N^{\frac{n}{2}}}{\sqrt{n!}}, \\
\label{alphaS} \alpha^{(S)}_{n}&=&\frac{1}{\left(N+1\right)^{\frac{1}{4}}}\frac{\sqrt{n!}}{\frac{n}{2}!}\left(\frac{1}{2}\sqrt{\frac{N}{N+1}}\right)^{\frac{n}{2}},
\end{eqnarray}

\noindent where the superscripts $F, C, S$ denote the Fock, coherent, and squeezed state, respectively. We underline that $N$ represents exactly the 
number of photons for the Fock state, while it is the average number $N$ of photons for the other two states where fluctuations in the photon numbers are present. Moreover, the probability amplitudes $\alpha_{n}^{(S)}$ are valid only for even values of $n$ and are zero otherwise.

It is worth mentioning that the QB+charger is a closed system. Interactions with possible external environments can lead to energy relaxation of the TLS and the loss of photons in the cavity. These processes can be characterized in terms of typical time scales $t_{r}$ and $t_{\gamma}$ respectively~\cite{Haroche_Book, Paladino08, Wendin17, Carrega20}. According to this we need to restrict our analysis to times, such that $t \ll t_{r}, t_{\gamma}$, in such a way to safely neglect the dissipation effects. However, we note that these constraints are comparable with the ones that are required by quantum information protocols~\cite{Devoret13}. Moreover, a protocol that is able to overcome the detrimental effects of energy relaxation in TLS have been theoretically discussed very recently in Ref. \cite{Bai20}.

\section{Figures of Merit of the Quantum Battery}\label{sec:characterization}

We now  characterize the performances of a two-photon QB comparing the three cavity states that were introduced in the previous Section.

In order to investigate the performance of the QB, it is necessary to study the time evolution of the initial state in Equation~(\ref{initial}). 
We observe that, over the basis given by the states $|g\rangle \otimes |n \rangle$ and $|e\rangle \otimes |n-2 \rangle$ ($n\geq 2$), with $|e\rangle$ excited state of the TLS, the Hamiltonian in Equation~(\ref{Htot}) assumes the simple form 

\begin{equation}
    \hat{\mathcal{H}}^{(n)}=\begin{pmatrix}\dfrac{\omega_a}{2}(n-1)& \lambda\sqrt{n(n-1)}\\\\
    \lambda\sqrt{n(n-1)}& \dfrac{\omega_a}{2}(n-1)  \end{pmatrix}
\end{equation}

\noindent where the notation keeps track of the number of photons $n$ and 
the dynamics remains confined in a two-dimensional space. This $2 \times 2$ matrix can be diagonalized in terms of the eigenstates 

\begin{equation}
|\psi^{(n)}_\pm\rangle=\frac{|g\rangle \otimes |n\rangle\pm|e\rangle \otimes |n-2\rangle}{\sqrt{2}},
\end{equation} 
with eigenvalues 
\begin{equation}
 E^{(n)}_\pm=\frac{\omega_a}{2} (n-1)\pm \lambda\sqrt{n(n-1)}.
 \end{equation}

In terms of these states, one can rewrite Equation~(\ref{initial}) as 
\begin{equation}\label{initial2}
   |\psi(0)\rangle=\sum_n \alpha_{n}\left(\frac{|\psi^{(n)}_+\rangle +|\psi^{(n)}_-\rangle}{\sqrt{2}}\right),
\end{equation}
that evolves in time as 
\begin{equation}
        |\psi(t)\rangle=e^{-i\hat{\mathcal{H}}t}|\psi(0)\rangle= \sum_n \alpha_n e^{-i\frac{\omega_a}{2}(n-1)t}\left(\frac{e^{+i\lambda\sqrt{n(n-1)}t}|\psi_{-}^{(n)}\rangle +e^{-i\lambda\sqrt{n(n-1)}t}|\psi_{+}^{(n)}\rangle}{\sqrt{2}}\right).
        \label{state_time}
\end{equation}

It is instructive to also consider the time dependent behaviour of the reduced density matrix that was obtained by tracing out the photon degrees of freedom when we consider the initial state in Equation~(\ref{initial2})
\begin{eqnarray}
\hat{\rho}_{\rm TLS}(t)&=&\sum_n\langle n | \hat{\rho}(t) |n \rangle \nonumber \\
&=&\sum_n\bigg\{ \bigg[p_n\sin^2(\lambda\sqrt{n(n-1)}t)|e\rangle\langle e|+p_n\cos^2(\lambda \sqrt{n(n-1)}t)|g\rangle\langle g|\bigg] \nonumber \\
&+& e^{i\omega_a t}|g\rangle\langle e| \bigg[i\sqrt{p_{n+2}p_n}\sin(\lambda \sqrt{(n+1)(n+2)}t)\cos(\lambda \sqrt{n(n-1)}t)\bigg] \nonumber \\
&-& e^{-i\omega_a t}|e\rangle\langle g| \bigg[i\sqrt{p_{n+2}p_n}\sin(\lambda \sqrt{(n+1)(n+2)}t)\cos(\lambda \sqrt{n(n-1)}t)\bigg] 
\label{rho_TLS_general}
\end{eqnarray}
\noindent where $p_{n}=|\alpha_{n}|^{2}$ and
\begin{equation}
\hat{\rho}(t)=  |\psi(t)\rangle \langle \psi(t)|
\end{equation}

\noindent is the total density matrix at time $t$ with $|\psi(t)\rangle$ the state in Equation~(\ref{state_time}). Equation~(\ref{rho_TLS_general}) can be written in the conventional form using the Bloch vector. Being written in terms of the identity matrix and the Pauli matrices, it reads~\cite{Nielsen, Grifoni98} 
 
\begin{equation}
    \hat{\rho}_{\rm TLS}(t)=\frac{1}{2}\left[\mathbb{I}+u(t)\hat{\sigma}_x+v(t)\hat{\sigma}_{y}+w(t)\hat{\sigma}_z\right].
\end{equation}

Here, starting from Equation~(\ref{rho_TLS_general}) and recalling that
\begin{eqnarray}
\mathbb{I}&\equiv&|e\rangle\langle e|+ |g\rangle\langle g|\\ 
\hat{\sigma}_{x}&\equiv&|g\rangle\langle e|+ |e\rangle\langle g| \\
\hat{\sigma}_{y}&\equiv&i(|g\rangle\langle e|- |e\rangle\langle g|)\\ 
\hat{\sigma}_{z}&\equiv&|e\rangle\langle e|- |g\rangle\langle g|
\end{eqnarray}
we can obtain the expressions for the components $u(t)$, $v(t)$, and $w(t)$
\begin{eqnarray}
\label{u}       u(t)&=&-\sin{(\omega_{a}t)}\sum_n\sqrt{p_np_{n+2}}\sin(\lambda\sqrt{(n+1)(n+2)}t)\cos{(\lambda\sqrt{n(n-1)}t)},\\
\label{v}        v(t)&=&\cos{(\omega_{a}t)}\sum_n\sqrt{p_np_{n+2}}\sin(\lambda\sqrt{(n+1)(n+2)}t)\cos{(\lambda\sqrt{n(n-1)}t)},\\
\label{w}        w(t)&=&-\sum_n p_n\cos{(2\lambda\sqrt{n(n-1)}t)},
\end{eqnarray}

Notice that, for the simple case of a Fock state, where the probability amplitude $\alpha_n$ is given by Equation~(\ref{alphaF}), the above expression assumes the simple form  

\begin{equation} 
u(t)=v(t)=0 \quad\quad\quad w(t)=-\cos{(2\lambda\sqrt{N(N-1)}t)}. 
\end{equation}

Instead, the explicit form for $u(t)$, $v(t)$, and $w(t)$ for the coherent and squeezed states can be obtained by replacing the $\alpha_n$ given in Equations~(\ref{alphaC}) and (\ref{alphaS}) into Equations~(\ref{u})--(\ref{w}), and cannot be further simplified.


\subsection{Stored Energy and Average Charging Power}
The energy stored in the QB at time $t$ is given by~\cite{Campaioli17, Thermodynamics18, Ferraro18, Crescente20, Crescente20b}

\begin{equation}\label{en}
    E(t)=\big[\langle \psi(t) |\hat{\mathcal{H}}_{QB}|\psi(t)\rangle-\langle \psi(0)|\hat{\mathcal{H}}_{QB}|\psi(0)\rangle\big],
\end{equation}
with 
\begin{equation}\label{Hbattery}
\hat{\mathcal{H}}_{QB}=\frac{\omega_a}{2}\hat \sigma_z
\end{equation} 
the contribution to the total Hamiltonian in Eq. (\ref{Htot}) that is associated to the QB.

Inserting the time evolved state of the system [see Equation~(\ref{state_time})] into Equation~(\ref{en}) one has 
\begin{equation}
\label{E(t)}
        E(t)=\omega_a\sum_n p_n\sin^2{(\lambda\sqrt{n(n-1)}t)},
\end{equation}

An important task in the context of QB is to store the maximum amount of energy in the fastest time. In this perspective, we define the maximum of the stored energy as~\cite{Andolina18, Crescente20b}
\begin{equation}
E_{\rm max}= \underset{t}{\text{max}} [E(t)] \equiv E(t_E), 
\end{equation}
where $t_E$ is the time at which the maximum occurs. In the case of a Fock state ($\alpha^{(F)}_n=\delta_{n,N}$), it is possible to analytically find the value of $t^{(F)}_E$ imposing the condition

\begin{equation} 
\sin^2(\lambda\sqrt{n(n-1)}t)=1. 
\end{equation}

Consequently the value of the maxima of the energy occurs when
\begin{equation}
\label{tF} 
t_E^{(F)}=\dfrac{\bigg(k+\frac{1}{2}\bigg)\pi}{\lambda\sqrt{n(n-1)}}. 
\end{equation}

\noindent where $k\in \mathbb{Z}$ indicates which maximum is considered. Instead, in the following, for the coherent and squeezed state, the charging time $t_E$ is obtained numerically.

\begin{figure}[H]
\includegraphics[scale=0.65]{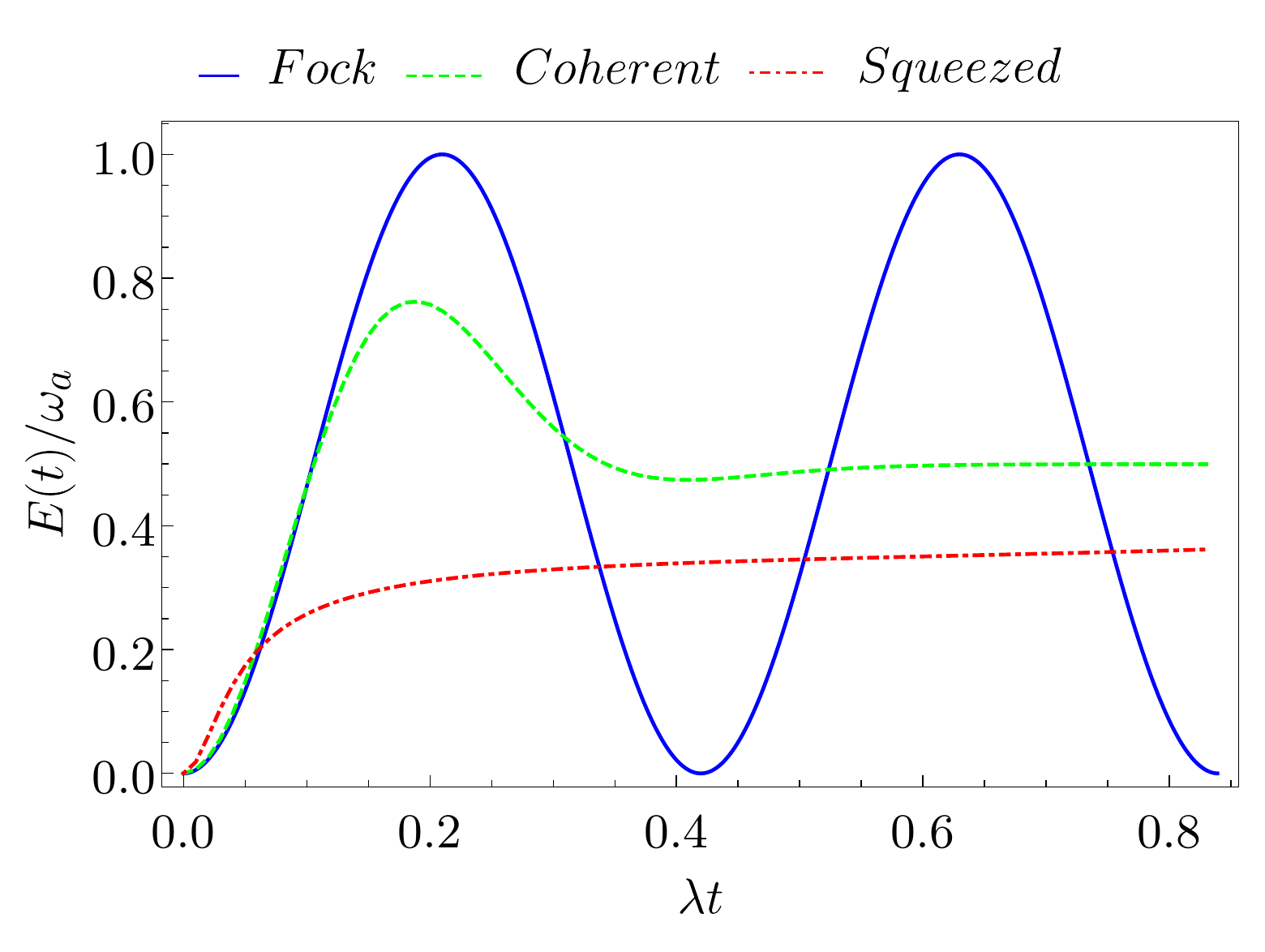}
\caption{(Color on-line) Behaviour of the stored energy $E(t)$ as a function of time for a Fock state (blue full curve), a coherent state (green dashed curve), and a squeezed state (red dash-dotted curve) with an average number of photons $N=8$.}
\label{Fig2}
\end{figure}

In  Figure~\ref{Fig2}, we report the behavior of the energy $E(t)$ (see Equation~(\ref{E(t)})) in units of $\omega_a$ as a function of time for the three different initial states of the cavity and for an average number of photons $N=8$. 

Here, one can see that a complete charging is only reached if the cavity is in a Fock state ($E^{(F)}_{\rm max}= \omega_{a}$). However, a coherent state with the same average number of photons ($N=8$ in the considered case) can also reach a quite large fraction of the maximal stored energy (the first maximum 
is $E^{(C)}_{\rm max}\approx 0.76\omega_{a}$). This aspect can be particularly relevant for the actual experimental realization of this kind of QB due to the fact that a coherent state is usually easier to realize in a cavity with respect to a well defined Fock state~\cite{Fink09}. Conversely, the squeezed state only reaches a very poor maximum charging $E^{(S)}_{\rm max}\approx 0.52\omega_{a}$.

Moreover, the minimal charging times for the Fock and coherent state are almost comparable and both are faster with respect to the squeezed state. Indeed, from Equation~(\ref{tF}), we have that $\lambda t_E^{(F)}\approx 0.21$, while $\lambda t_E^{(C)}\approx0.19$. Instead, the maximum of the energy of the squeezed state is achieved for the longer time $\lambda t_E^{(S)}\approx 1.53$.
It is also worth pointing out that, for a given quantum state of the charger, the charging times in the two-photon interaction model are usually shorter than the corresponding ones for a conventional dipolar single-photon coupling with the same average value of photons~\cite{Andolina18, Crescente20b, Ferraro18}. In fact, as reported in Ref.~\cite{Andolina18}, the energy for a single photon process is given by $E_{1ph}(t)= \sum_n p_n \sin^2(\lambda \sqrt{n}t)$, meaning that the time at which the maximum of the energy is reached scales as $1/\sqrt{n}$. Instead, in the two-photon case, it scales as $1/\sqrt{n(n-1)}$, leading to a charging time, which is $\sqrt{n-1}$ times faster fixing all other parameters.

From the previous discussion, one can infer that the Fock state is the best choice for the initial state of the charger, in order to store energy into a QB, in agreement with Ref.~\cite{Andolina18}. Conversely, the squeezed one is the less efficient, due to both the longer charging time and smaller maximum stored energy when compared to the other ones.
Notice that the Fock states still show the best charging performances in terms of stored energy and charging times also in comparison with other states of the general form in Equation (\ref{initial}).

\begin{figure}[H]
\includegraphics[scale=0.23]{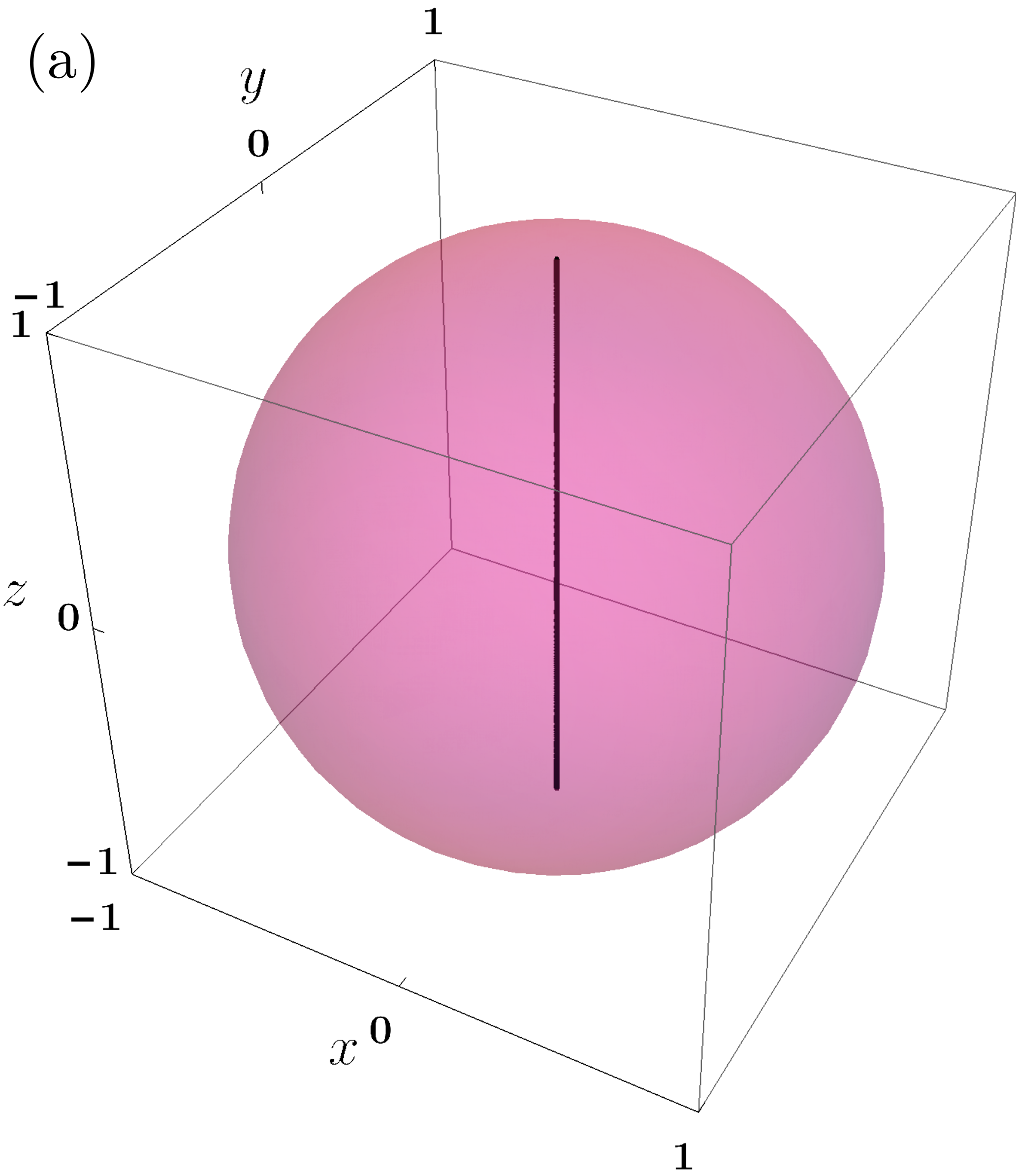}
\includegraphics[scale=0.23]{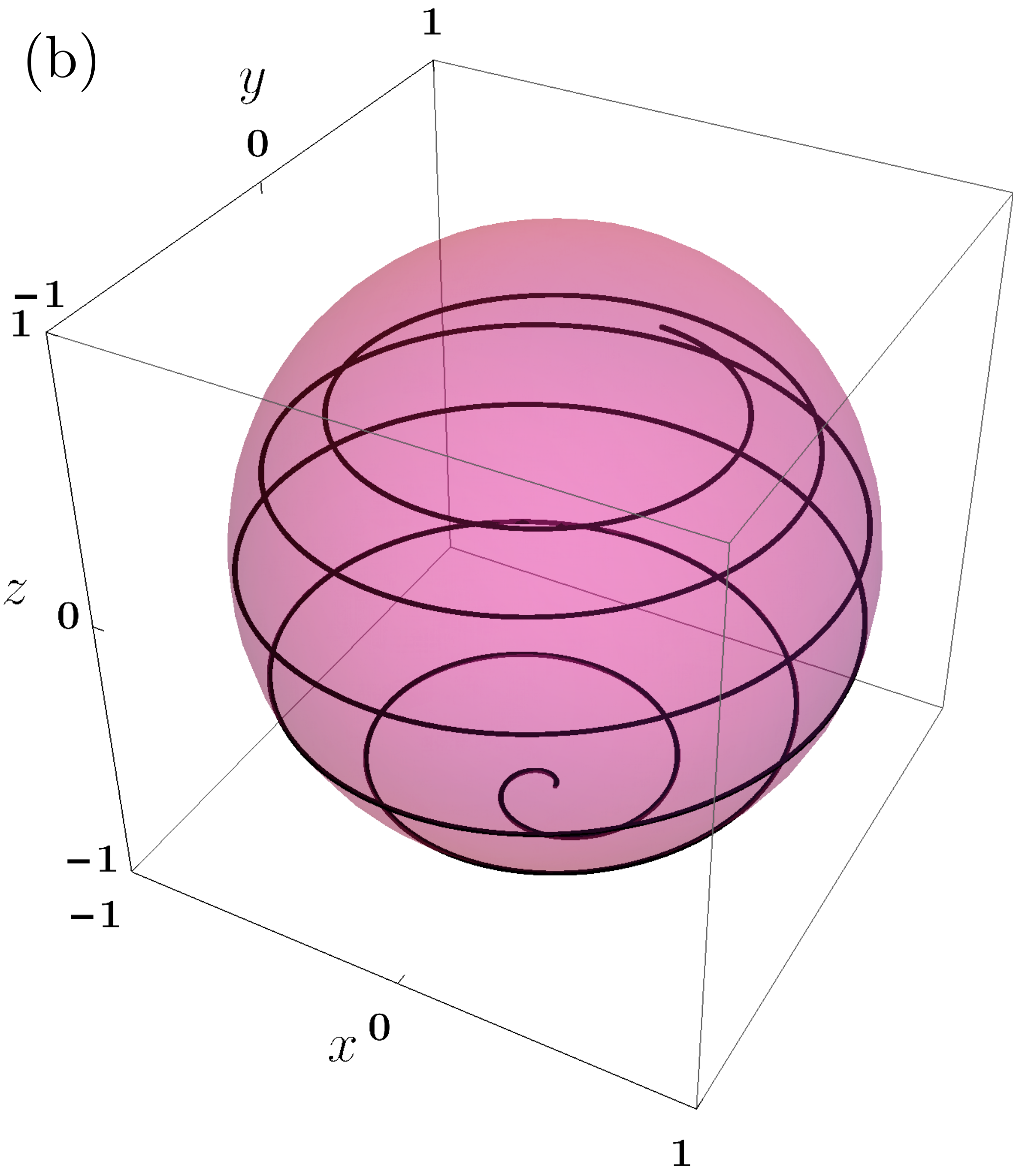}
\includegraphics[scale=0.23]{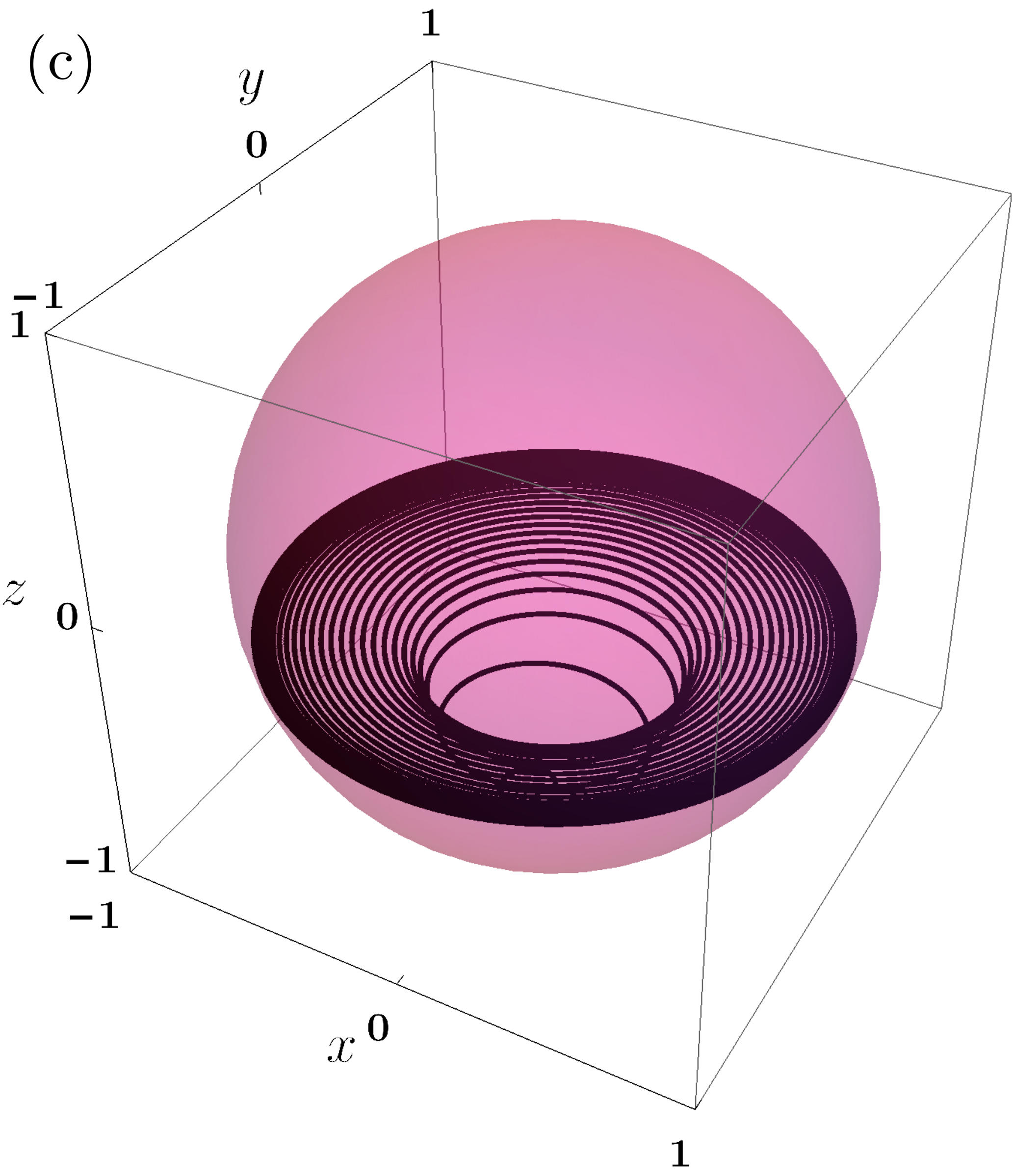}
\caption{(Color on-line) Time evolution of the quantum state of the TLS, up to the first maximum of the energy in Equation~(\ref{E(t)}), in the Bloch sphere for a Fock state (\textbf{a}), a coherent state (\textbf{b}), and a squeezed state (\textbf{c}) 
with an average number of photons $N=8$. The other parameters are: $\omega_{a}/\lambda=200$.}
\label{Fig3}
\end{figure}

The charging behaviour can be better understood by looking at the evolution of the TLS state on the Bloch sphere. In Figure~\ref{Fig3}, we show the path that is followed by the TLS state on the Bloch sphere for the three different 
cavity states, up to the first maximum of the stored energy. Notice that, here, the $|g\rangle$ and the $|e\rangle$ states are respectively represented by the south and north pole.

As discussed above, we initialize the TLS in the ground state. Here, we can observe that, only for the Fock state, the TLS state reaches the excited state, corresponding to the complete charging of the QB, while, for the coherent and the squeezed states, this never happens and the path of the state vector on the Bloch sphere is much more complicated. It is worth pointing out the fact that more involved initial states for the TLS, such as coherent superpositions of ground and excited state, are also characterized 
by complicated evolution in the Bloch sphere.

Another relevant figure of merit is the average charging power, which is defined as~\cite{Binder15, Campaioli17, Thermodynamics18, Ferraro18}

\begin{equation}
\label{P(t)}
    P(t)=\frac{E(t)}{t}.
\end{equation}

Regarding the energy, also in this case we are interested in achieving the maximum value of the charging power in the fastest possible time. Therefore, one needs to consider the maximum charging power~\cite{Andolina18, Ferraro18, Crescente20b}

\begin{equation}
P_{\rm max}= \underset{t}{\text{max}}\left[ P(t)\right]\equiv P(t_P),
\end{equation}
where $t_P$ represents the time at which the maximum occurs.

\begin{figure}[H]
\includegraphics[scale=0.65]{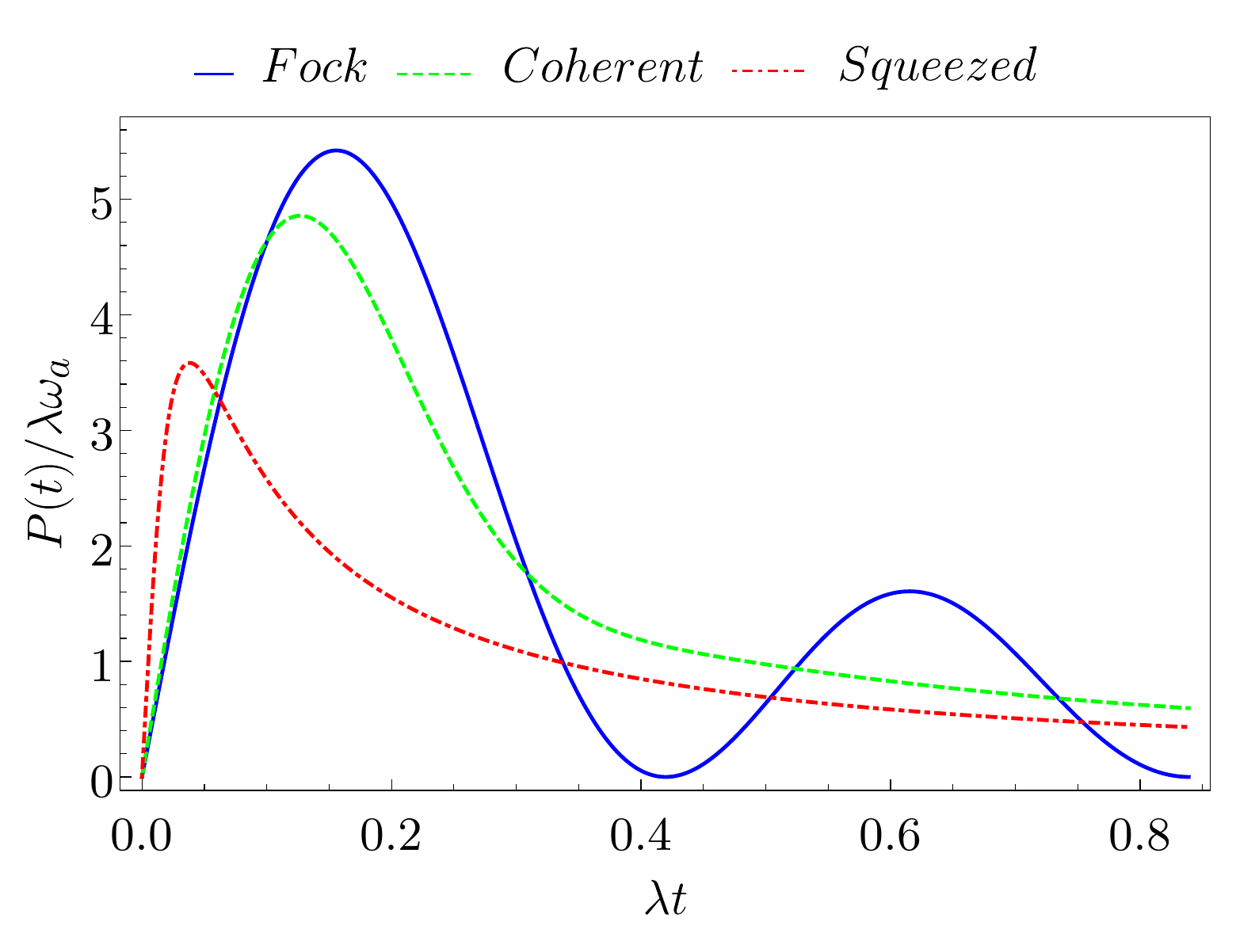}
\caption{(Color on-line) Behaviour of the average charging power $P(t)$ as a function of time for a Fock state (blue full curve), a coherent state 
(green dashed curve), and a squeezed state (red dotted-dashed curve) with an average number of photons $N=8$.}
\label{Fig4}
\end{figure}

In  Figure~\ref{Fig4}, we report the behaviour of $P(t)$ in Equation~(\ref{P(t)}) as a function of time for the three considered initial states of the cavity.  
Additionally, in this case, one finds that this quantity is maximal for the Fock state ($P_{\rm max}^{(F)}\approx 5.42 \lambda\omega_a$), while progressively decreasing for the coherent state ($P_{\rm max}^{(C)}\approx 4.86 \lambda\omega_a$) and the squeezed state ($P_{\rm max}^{(S)}\approx 3.58\lambda \omega_a$), respectively. However, the two former curves are quite similar in terms of the achieved maximum value and the corresponding times at which it occurs are very close ($\lambda t_P^{(F)} \approx 0.16$ and 
$\lambda t_P^{(C)} \approx 0.13$).
Different is the situation for the squeezed state that, in spite of the fact that the maximal averaged charging power is reached in a shorter time 
$\lambda t_P^{(S)}\approx 0.04$, shows a smaller value of the averaged charging power, as a consequence of the smaller stored energy.

\subsection{Energy Fluctuations}

\textls[-15]{In order to have a complete characterization of the QB, we now evaluate the quantum fluctuations that are associated to the stored energy, since they can induce detrimental effects on the charging performances~\cite{Crescente20, Crescente20b, Friis18}. We discuss the stability of the charging process in terms of the fluctuations of the stored energy at equal times, as represented by the correlator~\cite{Crescente20, Crescente20b, Friis18}}
\end{paracol}
\nointerlineskip

\begin{eqnarray}
\Xi(t)&=&\sqrt{\langle \psi(0)|\left[\hat{\mathcal{H}}_{QB}(t)-\hat{\mathcal{H}}_{QB}(0)\right]^2 |\psi(0)\rangle-\left[\langle \psi(0)|\left(\hat{\mathcal{H}}_{QB}(t)-\hat{\mathcal{H}}_{QB}(0)\right)|\psi(0)\rangle\right]^2} \nonumber \\
\label{xi} &=&\sqrt{E(t)\left[\omega_a-E(t)\right]},
\end{eqnarray}

\begin{paracol}{2}
\switchcolumn

\noindent with $|\psi(0)\rangle$ the initial state in Equation~(\ref{initial2}) and $\hat{\mathcal{H}}_{QB}(t)$ the TLS Hamiltonian evolved in time in the Heisenberg representation according to $\hat{\mathcal{H}}$. 

\begin{figure}[H]
\includegraphics[scale=0.65]{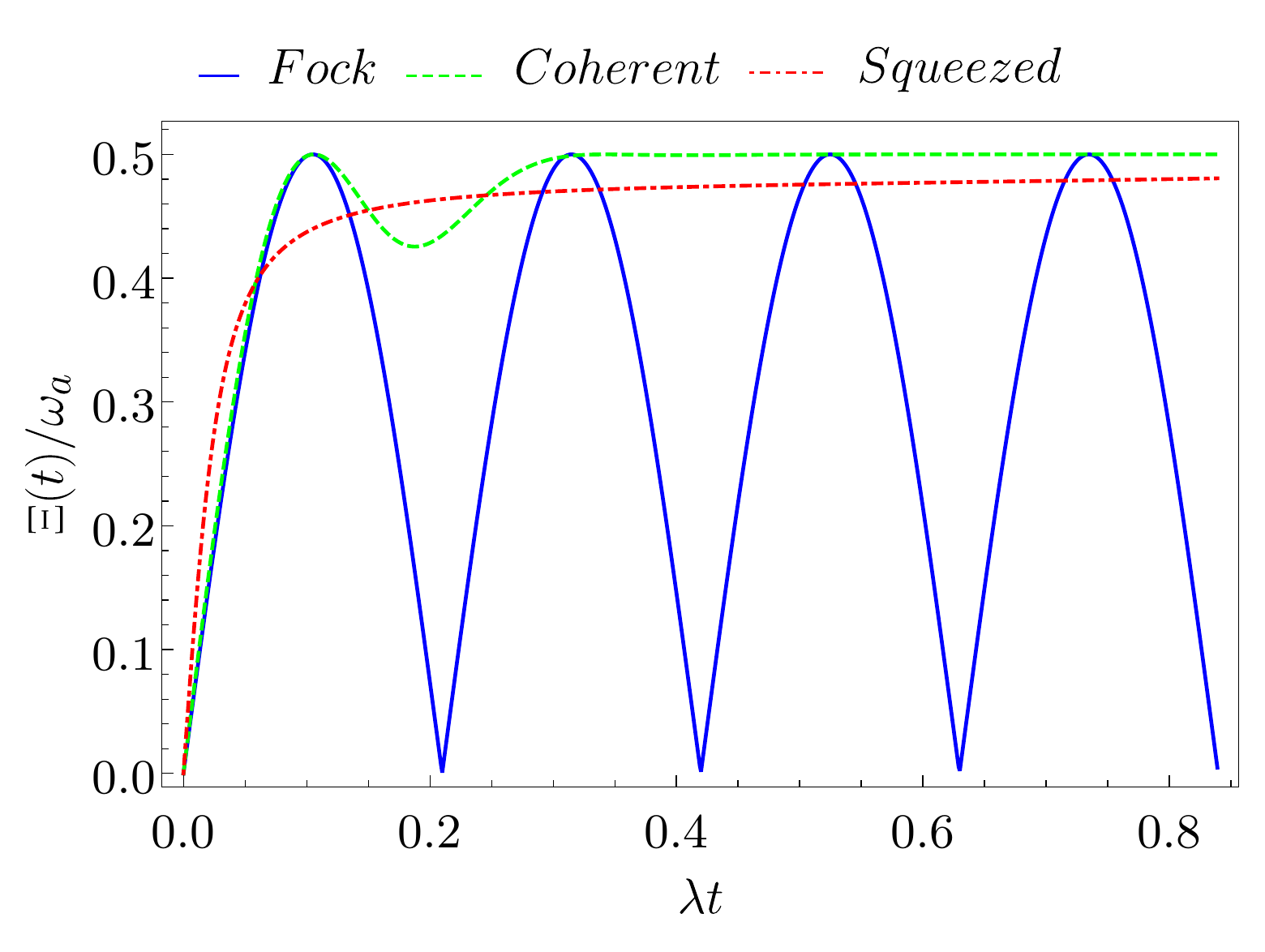}
\caption{(Color on-line) Behaviour of the stored energy fluctuations $\Xi(t)$ as a function of time for a Fock state (blue full curve), a coherent 
state (green dashed curve), and a squeezed state (red dotted-dashed curve) 
with an average number of photons $N=8$.}
\label{Fig5}
\end{figure}

In  Figure~\ref{Fig5}, we show the evolution of $\Xi(t)$ for the three initial states of the cavity. In the following, we are interested in studying 
the value of the correlator at time $t_E$, where the stored energy has its maximum, to understand how it affects the functionality of the QB. We then define

\begin{equation}
\Xi(t_E)\equiv \bar \Xi. 
\end{equation} 

Here, it clearly emerges that the Fock state has no energy fluctuations in correspondence of the maximum of the charging energy ($\bar \Xi^{(F)}=0$), as also reported in Ref.~\cite{Crescente20b}. This can be understood from Equation~(\ref{xi}) and it is a consequence of the 
fact that, at time $t_E$, the quantum state of the system (QB+charger) is 
separable ($|\psi(t_{E})\rangle =|e\rangle \otimes  |n-2\rangle$). This 
is not true for both the coherent and the squeezed states that show relevant fluctuations in the stored energy. 
Because neither of these states achieve $E_{\rm max}=\omega_a$, we will never observe $\bar \Xi=0$. In fact, for the coherent state, where the maximum energy was higher, we obtain $\bar \Xi^{(C)}\approx 0.43 \omega_a$, while, for the squeezed state, which is even worse in terms of energy storage, we have $\bar \Xi^{(S)} \approx 0.49\omega_a$.

Again, this confirms that the Fock state appears as the most convenient cavity state to build a good QB. Indeed, this particular initial state 
is not subject to stored energy fluctuations.


\subsection{Ergotropy}

Another relevant quantity to look at to determine the efficiency of a QB is  the so-called ergotropy~\cite{Allahverdyan04, Alicki13}. It 
consists in the maximal stored energy that can be converted into usable work. It can be extracted from the QB at a given time $t$ of its evolution (charging). In general, this quantity is different from the stored energy due to the fact that part of the energy may be locked into correlations and, therefore, cannot be extracted for further purposes~\cite{Francica20}. A general derivation of this quantity starts from the Hamiltonian of the QB, written as~\cite{Alicki13}
 
\begin{equation}\label{Hqbs}
\hat{\mathcal{H}}_{QB}=\sum_n\epsilon_n |\epsilon_n\rangle \langle \epsilon_n|
\end{equation}
with the energy eigenvalues ordered, such that $\epsilon_n<\epsilon_{n+1}$ and 
with $|\epsilon_n\rangle$ associated eigenvectors, as well as the density 
matrix at a given time 
\begin{equation}\label{rhotls}
\hat{\rho}_{\rm TLS} (t) =\sum_n r_n(t)|r_n(t)\rangle \langle r_n(t)|
\end{equation}
ordered, such that $r_n>r_{n+1}$ and with $|r_n\rangle$ eigenvectors. According to this, the work done on the system after a time $t$ is given by

\begin{equation} 
W(t)={\rm{Tr}}\{\rho_{\rm TLS}(t) \hat{\mathcal{H}}_{QB}\}-{\rm{Tr}}\{\rho_{\rm TLS}(0) \hat{\mathcal{H}}_{QB}\}. 
\end{equation}

The maximum work that can be extracted from the battery after a time $t$, called ergotropy, is defined as ${\cal E}= {\rm max}(-W)$.
Exploiting Equations~(\ref{Hqbs}) and (\ref{rhotls}), the ergotropy can be written as~\cite{Allahverdyan04} 

\begin{equation}\label{ergo}
\mathcal{E}(t)=\sum_{j,k}r_j(t)\epsilon_k\left(|\langle r_j(t)|\epsilon_k \rangle|^{2}-\delta_{jk}\right).
\end{equation}
It is worth pointing out that ${\cal E}=0$ when the initial state of the system is passive \cite{Alicki13}. Moreover, in general, the maximum extractable work ${\cal E}_{th}$ 
can be obtained when the final state of the system is thermal, i. e. $\rho_{\rm TLS}(t)=e^{-\beta \hat{\mathcal{H}}_{QB}}/Z$ (with $\beta$ the inverse temperature of the system and $Z={\rm {Tr}}\{e^{-\beta \hat{\mathcal{H}}_{QB}}\}$). Subsequently, the bound on the maximum and minimum extractable 
work is given by $0\leq {\cal E}\leq {\cal E}_{th}$~\cite{Allahverdyan04}.

\begin{figure}[H]
\includegraphics[scale=0.6]{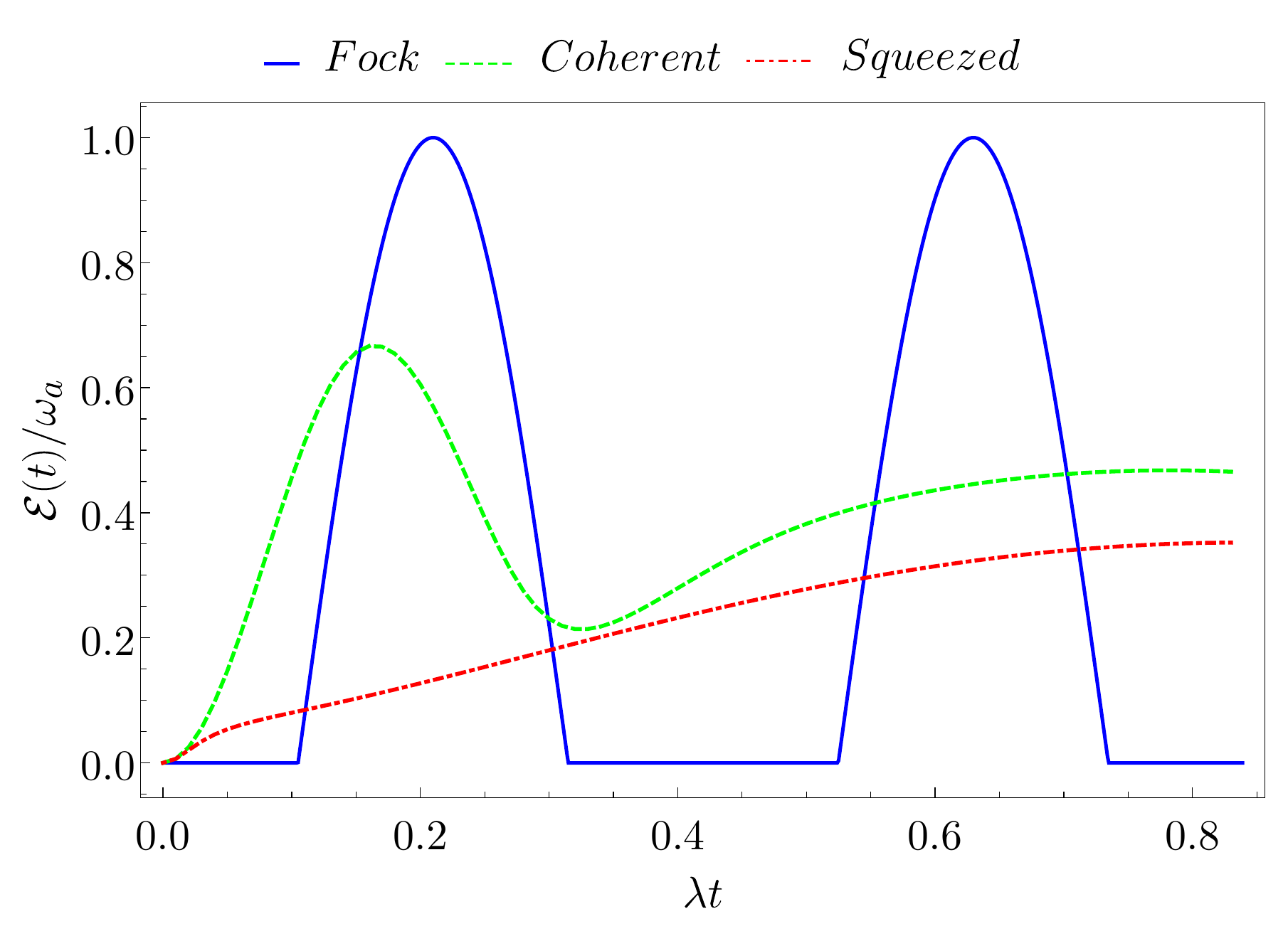}
\caption{(Color on-line) Behaviour of the ergotropy $\mathcal{E}(t)$ as a function of time for a Fock state (blue full curve), a coherent state (green dashed curve), and a squeezed state (red dash-dotted curve) with average number of photons $N=8$.}
\label{Fig6}
\end{figure}

We now want to find the explicit form of the ergotropy for our system. To do so, we need to diagonalize the density matrix of the TLS in Equation~(\ref{rho_TLS_general}). Its eigenvalues are

\begin{equation} 
r_s(t)=\frac{1+(-1)^s\sqrt{1-4\det\hat{\rho}_{\rm TLS}(t)}}{2}, 
\end{equation}

\noindent where $s=0,1$. Moreover, the eigenvalues of $\hat{\mathcal{H}}_{QB}$ in Equation~(\ref{Hbattery}) are $\varepsilon_s=(-1)^s\omega_a/2$.
Consequently, starting from Equation~(\ref{ergo}), the ergotropy at time $t$ can be written as
\begin{equation}
\label{erg} 
\mathcal{E}(t)=E(t)-\sum_{s=0,1}r_s(t)\varepsilon_s=E(t)-\frac{\omega_a}{2}\left[1-\sqrt{1-4\det\hat{\rho}_{\rm TLS}(t)}\right].
\end{equation}

Figure~\ref{Fig6} shows its time-evolution.  Here, one has a qualitatively similar behaviour with respect to the one that is observed for the stored energy. We observe that only the Fock state reaches the maximum of the ergotropy $\mathcal{E}_{max}=\omega_a$. This occurs at a time $t^{(F)}_{E}$, where the energy has its maxima and the state of the TLS is pure, confirming the relevant performances of this cavity state.

To better quantify the actual fraction of extractable energy, we now evaluate the ratio between the ergotropy and stored energy, as given by
\begin{equation}
\eta(t) \equiv \frac{\mathcal{E}(t)}{E(t)}=1-\frac{\omega_{a}}{2E(t)}\left[1-\sqrt{1-4\det{\hat{\rho}_{TLS}(t)}}\right].
\end{equation}

In Figure~\ref{Fig7}, we compare the behavior of the ratio $\eta$ for the three considered initial states as a function of time. As one can observe, at quite short times the coherent state allows an almost complete extraction of the energy as usable work. However, in this region, the energy that is stored in the system can be quite small [see Figure \ref{Fig2}]. For the Fock state it is possible to completely extract the stored energy only for narrow time windows in correspondence of $t^{(F)}_{E}$, with the advantage that the stored energy approaches or is equal to the maximum value $E_{\rm max}^{(F)}=\omega_{a}$. Finally, the squeezed state reaches (together with the coherent state) a unitary value of the ratio only at long enough times, where again the energy stored in the QB is very limited and possible dissipative effects could came into play.

 \begin{figure}[H]
\includegraphics[scale=0.6]{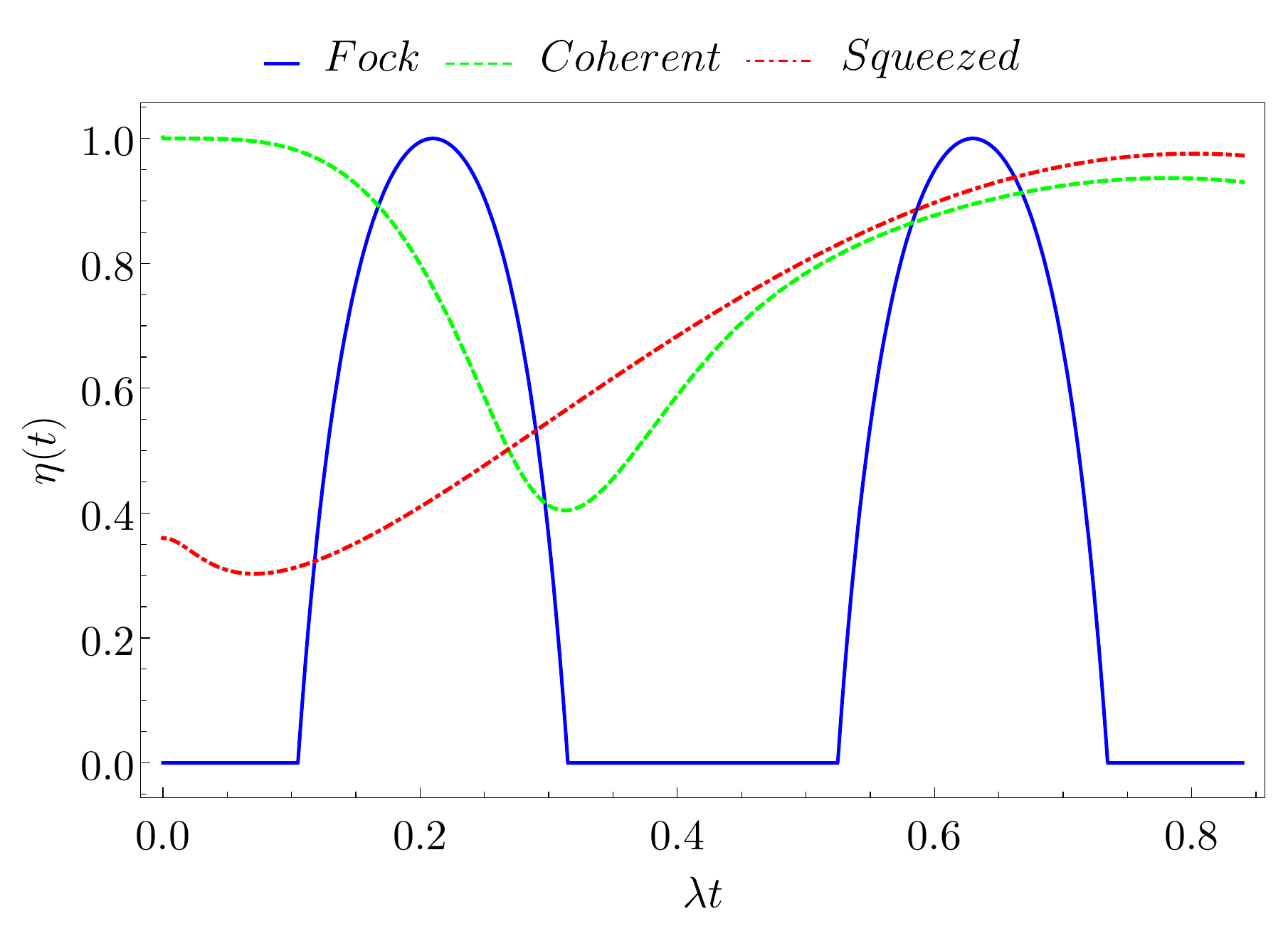}
\caption{(Color on-line) Behaviour of the ratio $\eta(t)$ as a function of time for a Fock state (blue full curve), a coherent state (green dashed curve), and a squeezed state (red dash-dotted curve) with average number of photons $N=8$.}
\label{Fig7}
\end{figure}

This result is a consequence of the nature of the three considered states. In fact, as also stated in Ref.~\cite{Andolina19}, the more there is a mixing between the TLS and the photonic part, the less energy we can extract, since the entanglement between the QB and charger has a negative impact on the possibility to extract work.

\section{Conclusions \label{sec:conclusions}}

We have characterized various figures of merit for a quantum battery given by a qubit, namely a two-level system, coupled with a cavity radiation through a two-photon coupling (quadratic in the quantum electromagnetic field). As possible initial conditions for the quantum radiation in the cavity, we have inspected a Fock state, a coherent state, and squeezed state. According to our analysis, the Fock state emerges as the ideal candidate for optimizing the performances of the quantum battery, being the only state able to reach a complete charging of the quantum battery, in short times and without showing fluctuations. Quite remarkably, this kind of state also allows for completely extracting its maximal stored energy. A coherent state with the same average number of photons also shows quite interesting performances, in particular for what concerns the fraction of extractable work at short enough times, even if it is affected by fluctuations in the stored energy. This is not true for a squeezed state whose performances are very poor, which makes it the worst state for implementing a quantum battery. Indeed, it can only store a fraction of the optimal energy and it is affected by very strong fluctuations. The present analysis gives important hints towards the possible implementations of quantum batteries coupled to cavity radiation.

\vspace{6pt}

\authorcontributions{Conceptualization, A.D., A.C., M.C., and D.F.; formal analysis, A.D., A.C.,
and D.F.; writing-original draft preparation, D.F.; writing-review and editing, A.D.,
A.C., M.C., and M.S.; supervision, M.S. All authors have read and agreed to the published
version of the manuscript.}

\funding{This research received no external funding.}




\dataavailability{The data and analysis used in this work are available from the corresponding
author upon reasonable request.} 
\newpage
\conflictsofinterest{The authors declare no conflict of interest.} 

\abbreviations{The following abbreviations are used in this manuscript:\\\noindent 
\begin{tabular}{@{}ll}
QB & Quantum Battery\\
TLS & Two-Level-System\\
\end{tabular}}


\end{paracol}

\reftitle{References}


\end{document}